\documentclass[usegraphicx, useAMS, usenatbib]{mn2e}
\usepackage{times}

\newcommand{\apj}{ApJ}
\newcommand{\aj}{AJ}
\newcommand{\aap}{A\&A}
\newcommand{\mnras}{MNRAS}
\newcommand{\pasp}{PASP}
\newcommand{\an}{Astr. Nachr.}
\newcommand{\apjs}{ApJS}
\newcommand{\kms}{km~s$^{-1}$}

\title[Binarity and multiperiodicity in HADS variables]{Binarity and
multiperiodicity in high-amplitude $\delta$ Scuti stars}

\author[Derekas et al.]{A. Derekas$^{1}$\thanks{E-mail:
derekas@physics.usyd.edu.au}, L. L. Kiss$^1$, T. R. Bedding$^{1}$, M. C. B.
Ashley$^{2}$, B. Cs\'ak$^{3,4}$, A. Danos$^1$, 
\newauthor J. M. Fernandez$^{5}$, G. F\H ur\'esz$^{5}$, Sz. M\'esz\'aros$^{3}$,
Gy. M. Szab\'o$^{6}$, R. Szak\'ats$^{6}$,  P. Sz\'ekely$^{6}$, 
\newauthor K. Szatm\'ary$^{6}$\\ 
\\$^1$Sydney Institute For Astronomy, School of Physics, University of Sydney, NSW 2006,
Australia
\\$^2$School of Physics, Department of Astrophysics and Optics, University of 
New South Wales, Sydney, NSW 2052, Australia 
\\$^3$Department of Optics and Quantum Electronics, University of Szeged, 
Hungary 
\\$^4$Max Planck Institute for Astronomy, K\"onigstuhl 17, D-69117 Heidelberg,
Germany
\\$^5$Harvard-Smithsonian Center for Astrophysics, 60 Garden Street, 
Cambridge, MA 02138, USA  
\\$^6$Department of Experimental Physics and Astronomical Observatory, 
University of Szeged, Szeged, D\'om t\'er 9, 6720 Hungary}

\begin{document}

\date{Accepted ... Received ..; in original form ..}


\maketitle

\begin{abstract}

We have carried out a photometric and spectroscopic survey of bright
high-amplitude $\delta$~Scuti (HADS) stars. The aim was to detect binarity and
multiperiodicity (or both) in order to explore the possibility of combining
binary star astrophysics with stellar oscillations. Here we present the first
results for ten, predominantly southern, HADS variables. We detected the
orbital motion of RS~Gru with a semi-amplitude of $\sim$6.5~\kms\ and 11.5~days
period. The companion is inferred to be a low-mass dwarf star in a close orbit
around RS~Gru. We found multiperiodicity in RY~Lep both from photometric and
radial velocity data and detected orbital motion in the radial velocities with
hints of a possible period of 500--700~days.  The data also revealed that
the amplitude of the secondary frequency is variable on the time-scale of a few
years, whereas the dominant mode is stable. Radial velocities of AD~CMi
revealed cycle-to-cycle variations which might be due to non-radial pulsations.
We confirmed the multiperiodic nature of BQ~Ind, while we obtained the first
radial velocity curves of ZZ~Mic and BE Lyn. The radial velocity curve and the
O--C diagram of CY~Aqr are  consistent with the long-period binary hypothesis.
We took new time series photometry on XX~Cyg, DY~Her and DY~Peg, with which we
updated their O--C diagrams.

\end{abstract}


\section{Introduction}

$\delta$~Scuti stars are short-period pulsating variables of A-F spectral
types, located at the intersection of the main sequence and the classical
instability strip in the Hertzsprung-Russell diagram. Typical periods are in
the order of a few hours with amplitudes less than 1~mag. A prominent group
within the family comprises the high-amplitude $\delta$~Scuti stars (HADS),
which have V-band amplitudes larger than 0.3~mag. Population~II members of the
group are also known as SX~Phoenicis stars, often found in globular clusters
(for a review of $\delta$~Scuti stars see e.g. \citet{rod01}).

HADS's are the short-period counterparts of the classical Cepheids, excited by
the $\kappa$-mechanism and pulsating in one or two radial modes, usually in the
fundamental and first-overtone modes \citep{mcn00}. The data in \citet{rod01}
show that there is no rapidly rotating HADS ($v\sin{i}\leq40$~\kms), suggesting
an intimate relationship between the rotational state and the excitation of
pulsations. In recent years, the number of HADS known with multimode
oscillations has rapidly grown, hinting a new potential for asteroseismic
studies of these objects \citep[e.g.][]{por03,por05}. Several investigations
suggested that some of the stars may have non-radial pulsation modes present
\citep{mcn00,por05}. However, an important parameter in modeling stellar
oscillations is the mass of the star, which is usually constrained from
evolutionary models. This will inevitably lead to great uncertainties in any
kind of modeling attempts, so that independent mass estimates could be of
paramount importance. Thus, binary $\delta$~Scutis may play a key role in
understanding oscillations of these stars.

There has been a great interest recently in $\delta$~Scuti stars that reside in
eclipsing binary systems \citep[e.g.][]{kim03,mkr06,soy06,pig07,chr07}.
Currently, we know about 40 such $\delta$~Scutis, of which only one belongs to
the HADS group \citep{chr07}. There are also a handful of non-eclipsing binary
HADS \citep{rod01}, usually deduced from the apparent cyclic period changes but
with a few exceptions, like the single-lined spectroscopic binary SZ~Lyn. The
low number of binary $\delta$~Scutis suggests a strong observational bias,
because one needs accurate observations with a long time span to detect
multiplicity unambiguously \citep{rod01}.

In this paper we present the results of our investigations into binarity and
multiple periodicity in bright HADS variables. The sample was initially
selected from the variable star catalogue of the Hipparcos satellite, which
contains 21 ``SX~Phe'' type stars \citep{per97b}. Of these, here we discuss 9
stars (and RY~Lep in addition), i.e. our present sample contains almost half of
all known bright HADS. Using a wide range of telescopes and instruments, we
have been monitoring the target stars over the last five years, extending the
earlier studies by our group \citep{kis95,kis02,der03,sza08}.

The paper is organized as follows. The observations and the data analysis are
described in Sect. 2. The main discussion is in Sect. 3, in which the results
for individual stars are presented. A brief summary is given in Sect. 4.

\section{Observations and data reduction}

The observed stars and their main observational properties are listed in Table\
\ref{obsprop}. The full log of observations is given in Table\ \ref{obslog} in
the Appendix. All data presented in this paper are available for download from
the CDS, Strasbourg. 

Observations were carried out using seven different instruments at five
observatories in Australia, Hungary and the USA on a total of 65 nights between
2003 October and 2008 July. In the following we briefly describe the telescopes
and the detectors used in this project, in an order of increasing mirror size:

\begin{table}   
\begin{center} 
\caption{\label{obsprop} The list of programme stars. The asterisks mark
multiperiodic stars, for which this Table contains the dominant period only.
 References for the parameters are the following: (a) \citet{rod96}; (b)
\citet{rod01}; (c) \citet{per97b}; (d) \citet{kho8588}; (e) \citet{sza08}; (f)
\citet{bla03}; (g) \citet{rod95b}; (h) \citet{fu03}; (i) \citet{der03}. }
\label{obsprop}
\setlength{\tabcolsep}{1.8mm}  
\begin{tabular}{|llrrlc|} 
\hline\hline  
Star & Pop.$^{\rm a}$ & V$_{\rm max}$ & V$_{\rm min}$ & P~(days) & Obs.\\
\hline  

RY~Lep* & I &  8$.\!\!^{\rm m}$20 & 9$.\!\!^{\rm m}$10 & 0.22514410$^{\rm c}$ &
$I/sp.$\\

AD~CMi* & I & 9$.\!\!^{\rm m}$21 & 9$.\!\!^{\rm m}$51 &  0.12297443$^{\rm d}$ & $sp.$\\ 

BE~Lyn &  I & 8$.\!\!^{\rm m}$60 & 9$.\!\!^{\rm m}$00 & 0.09586952$^{\rm e}$ & $sp.$\\

DY~Her & I & 10$.\!\!^{\rm m}$15 & 10$.\!\!^{\rm m}$66 &  0.14863135$^{\rm d}$ & $V$\\

XX~Cyg & II & 11$.\!\!^{\rm m}$28 & 12$.\!\!^{\rm m}$13 &  0.13486511$^{\rm f}$ & $V$\\

BQ~Ind* & II$^{\rm b}$ & 9$.\!\!^{\rm m}$78 & 10$.\!\!^{\rm m}$05 &
0.08200015$^{\rm c}$ & $I$\\

ZZ~Mic* &  I & 9$.\!\!^{\rm m}$27 & 9$.\!\!^{\rm m}$69 & 0.06718350$^{\rm d}$ &
$BV/sp.$\\  

RS~Gru & I & 7$.\!\!^{\rm m}$92 & 8$.\!\!^{\rm m}$51 & 0.14701131$^{\rm g}$ &
$BVI/sp.$\\ 

CY~Aqr & II & 10$.\!\!^{\rm m}$42 & 11$.\!\!^{\rm m}$16 & 0.06103833$^{\rm h}$ &
$BVI/sp.$\\   

DY Peg & II & 9$.\!\!^{\rm m}$95 & 10$.\!\!^{\rm m}$62 & 0.07292630$^{\rm i}$ & $V$\\

\hline\hline   
\end{tabular} 
\end{center}  
\end{table}

\begin{table}   
\begin{center}
\caption{\label{maxtimes} New times of maximum (HJD-2400000).}  
\label{maxtimes}  
\begin{tabular}{|lrclrc|}  
\hline\hline   
Star & {$\rm HJD_{max}$} & Filter & Star & {$\rm HJD_{max}$} & Filter \\ 
\hline

RS Gru & 52920.0196 & $V$ & CY Aqr & 53334.9453 & $I$  \\
RS Gru & 52921.9311 & $V$ & CY Aqr & 53336.9592 & $I$  \\
RS Gru & 52922.0772 & $V$ & CY Aqr & 53337.9357 & $I$  \\
RS Gru & 52923.9905 & $V$ & CY Aqr & 54307.5293 & $V$  \\
RS Gru & 52925.0188 & $V$ & CY Aqr & 54308.5073 & $V$  \\
CY Aqr & 52920.9223 & $V$ & XX Cyg & 54307.4294 & $V$  \\
CY Aqr & 52920.9827 & $V$ & XX Cyg & 54309.4515 & $V$  \\
CY Aqr & 52921.0439 & $V$ & XX Cyg & 54677.3633 & $V$  \\
CY Aqr & 52923.0587 & $V$ & DY Her & 54304.4772 & $V$  \\
CY Aqr & 52926.9643 & $V$ & DY Peg & 54305.4731 & $V$  \\
CY Aqr & 52927.0258 & $V$ &  && \\
\hline\hline
\end{tabular} 
\end{center}  
\end{table}

\begin{itemize}

\item[--] {\it Szeged Observatory, 0.4~m  (Sz40)}

$V$-band CCD photometry of XX~Cyg was carried out with the 0.4~m Newtonian
telescope at Szeged Observatory. The telescope has an
11$^{\prime}$$\times$17$^{\prime}$ field of view and the detector was an SBIG
ST-7 CCD camera (765 $\times$ 510~pixels at 9~$\mu$m). We took observations for
XX~Cyg with exposure time of 45~s. 

\item[--] {\it Siding Spring Observatory, 0.5~m  (APT50)}

Time-series CCD photometry was obtained for RY~Lep, BQ~Ind and CY~Aqr with the
Automated Patrol Telescope (APT) at Siding Spring Observatory, which is owned
and operated by the University of New South Wales (UNSW). The telescope was
originally a Baker-Nunn design converted into CCD imaging \citep{car92} and has
a three-element correcting lens and an f/1 spherical primary mirror. The camera
has  an EEV CCD05-20 chip with 770$\times$1150 (22.5~$\mu$m) pixel to image a
2$\times$3~deg$^2$ field of view. We obtained $I$-band images with exposure
times between 3~s and 60~s, depending on the star and the weather conditions.

\item[--] {\it Siding Spring Observatory, 24$^{\prime\prime}$/0.6~m (SSO60)}

Photoelectric photometry of ZZ~Mic, RS~Gru and CY~Aqr was obtained with the
0.6~m f/18 Cassegrain-reflector, on which a single-channel photometer was
mounted. It had a computer-controlled 8-hole filter wheel, for which the dwell
times on each filter can be varied. The detector was a thermoelectrically
cooled Hamamatsu R647-4 photomultiplier tube with a 9~mm diameter bi-alkali
(blue-sensitive) photocathode \citep{han00}. We used $B$, $V$ and $I$ filters
and the exposure time varied between 15~s and 30~s, depending on the brightness
of the observed star. 

\item[--] {\it  Piszk\'estet\H o Station, 0.6~m (P60)}

$V$-band CCD photometry was obtained on CY~Aqr, XX~Cyg, DY~Her and DY~Peg with
the 60/90/180~cm Schmidt-telescope mounted at the Piszk\'estet\H o Station of
the Konkoly Observatory. The detector was a Photometrics AT200 CCD camera
(1536$\times$1024 9~$\mu$m pixels, FOV=28$^\prime$$\times$19$^\prime$).

\item[--] {\it Fred Lawrence Whipple Observatory, 1.5~m (MH150)}

High resolution spectra were obtained on BE~Lyn with the Tillinghast Reflection
Echelle Spectrograph (TRES) and the 1.5~m telescope at the Fred Lawrence
Whipple Observatory on Mt.  Hopkins, Arizona. TRES is a high-throughput
fiber-fed echelle. It is cross-dispersed, yielding a passband of 380-920~nm
over the 51 spectral orders. It accommodates 3 optical fiber pairs
(science+sky) of different diameters, to offer a match for seeing conditions.
Simultaneous ThAr calibration is also available via a separate fiber. The
avaliable resolutions are 64K, 35K and 31K, depending on the fiber size
selected (1.5, 2.3 or 3.2~arcsec, respectively). The observations were taken as
part of the instrument commissioning, using the small fiber and 300~s
integration times. The 4.6k$\times$2k detector was binned 2$\times$2 in order
to improve the duty cycle (15~s readout time), providing a slightly
undersampled FWHM of 2.0~pixels.

\item[--] {\it Siding Spring Observatory, 2.3~m (SSO230)}

Spectroscopic observations were carried out for RY~Lep, AD~CMi, ZZ~Mic, RS~Gru
and CY~Aqr with the 2.3~m ANU telescope at the Siding Spring Observatory,
Australia. All spectra were taken with the Double Beam Spectrograph using the
1200~mm$^{-1}$ gratings in both arms of the spectrograph. The projected slit
width was 2$^{\prime\prime}$ on the sky, which was about the median seeing
during our observations. The spectra covered the wavelength ranges
4200--5200~\AA\ in the blue arm and 5700--6700~\AA\ in the red arm. The
dispersion was 0.55~\AA~px$^{-1}$, leading to a nominal resolution of about
1~\AA. The exposure time varied between 50~s and 180~s depending on the
observed star and the weather conditions.

\item[--] {\it Anglo-Australian Observatory, 3.9~m (AAT)}

We used the 3.9~m Anglo-Australian Telescope equipped with the UCLES
spectrograph for 4.2 hours of high-resolution spectroscopy of RS~Gru. The
observations were taken during service time. Our echelle spectra include 56
orders with a central wavelength of 6183~\AA\ and a resolving power
$\lambda/\Delta\lambda\approx$40 000. The exposure time was 300~s.

\end{itemize}

All data were reduced with standard tools and procedures. Photoelectric
photometry taken with the SSO60 instrument were transformed to the standard
system using the coefficients from \citet{ber04}.  Uncertainties in the
Fourier amplitudes were calculated following the considerations of
\citet{kje03}. The $BV$ standard magnitudes of comparison stars were taken
from \citet{kha01}, while $I$ standard magnitudes were taken from the DENIS
catalogue  \citep{den05}. The CCD observations were reduced in
IRAF\footnote{IRAF is distributed by the National Optical Astronomy
Observatories, which are operated by the Association of Universities for
Research in Astronomy, Inc., under cooperative agreement with the National
Science Foundation.}, including bias removal and flat-field correction
utilizing sky-flat images taken during the evening or morning twilight.
Differential magnitudes were calculated with aperture photometry using two
comparison stars of similar brightnesses.

Thanks to the dense sampling of the light curves, new times of maximum light
for monoperiodic stars were easy to determine from the individual cycles. This
was done by fitting fifth-order polynomials to the light curves around the
maxima. We estimate the typical uncertainty to be about $\pm$0.0003~d. The new
times of maximum light are listed in Table\ \ref{maxtimes}.

For the multiperiodic stars, we performed standard Fourier-analysis with
prewhitening using Period04 \citep{len05}. Least-squares fitting of the
parameters was also included and the $S/N$ of each frequency was calculated
following \citet{bre93}.

All spectra were reduced with standard tasks in IRAF. Reduction consisted of
bias and flat field corrections, aperture extraction, wavelength calibration,
and continuum normalization. We checked the consistency of wavelength
calibrations via the constant positions of strong telluric features, which
verified the stability of the system. Radial velocities were determined with
the task {\it fxcor}, applying the cross-correlation method using a
well-matching theoretical template spectrum from the extensive spectral library
of \citet{mun05}. For consistency, the velocities presented in this paper were
all determined from a 50~\AA\ region centered on the H$\alpha$ line. The
high-resolution spectra for BE Lyn and RS Gru allowed us to compare hydrogen
and metallic line velocities, which will be discussed in a subsequent paper. We
made barycentric corrections to every radial velocity value. Depending on the
signal-to-noise of the spectra, the estimated uncertainty of the radial
velocities ranged from 1--5~\kms. Radial velocities from the AAT and MH150
Echelle spectra have much better accuracy, $\leq$100~m~s$^{-1}$.

\section{Results}

\subsection{RS~Gruis}

\begin{figure}
\begin{center}
\includegraphics[width=8cm]{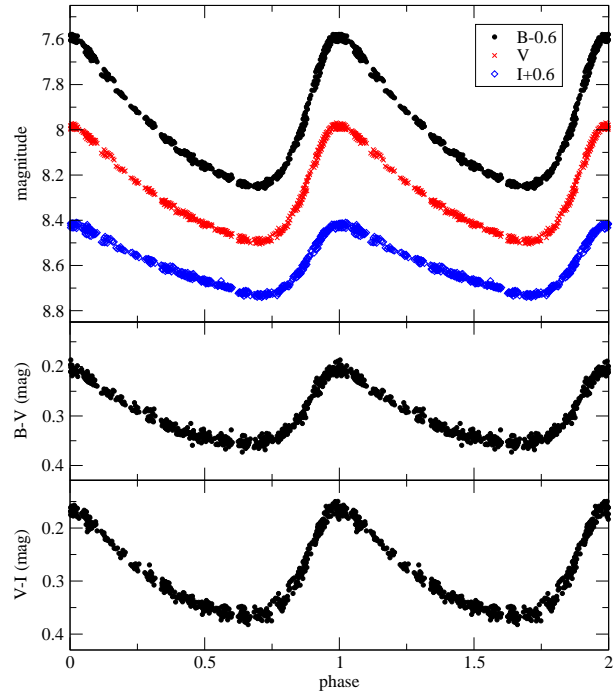}   
\end{center}
\caption[]{\label{rsgrufaz} Standard light and colour variations of RS Gru
(E$_{0}$=2452920.0196; P=0.14701131 d).} 
\end{figure}

RS~Gru (HD~206379; HIP~107231) is a monoperiodic HADS with a pulsation period
of 0.147~d and a mean magnitude of $\sim$7.9~mag. Its light variation was first
detected by \citet{hof56} and studied later by \citet{egg56} and \citet{oos66}.
\citet{kin61} took photometric and spectroscopic observations and measured a
mean velocity of 81~\kms\ with a velocity amplitude of $\sim$45~\kms.
\citet{mnf76} obtained {\it uvby$\beta$} photometry and spectrographic data and
determined physical parameters. \citet{cit76} acquired photoelectric radial
velocity curves on two nights, while further photometric observations were
taken by \citet{dea77}. New radial velocity measurements taken by
\citet{bal78a} showed unambiguously the variation of the center-of-mass
velocity, indicating the binary nature of RS~Gru but the orbital period was not
determined for nearly three decades. Further investigations were done by
\citet{bre80,and83,mcn85,ant86,gar90,cla90,rod90b}. Period decrease was found
by \citet{rod95a,rod95b} and physical parameters were also calculated by
\citet{rod95b}. \citet{jon04} took high-quality spectroscopic measurements and
determined the radius and the absolute magnitude for RS~Gru. They again showed
unambiguously that RS~Gru is a spectroscopic binary with an orbital period of
approximately two weeks but no exact period was given.

We obtained standard $BVI$ photoelectric photometry using SSO60 on 4 nights in
2003 and 1 night in 2004. For differential magnitudes we used two comparison
stars: comp=HD~207193 ($V=6.79$~mag, $B-V=0.35$~mag) and check=HD~207615
($V=8.53$~mag, $B-V=0.08$~mag). The full log of observations is given in Table\
\ref{obslog} and the light and colour variation are plotted in Fig.\
\ref{rsgrufaz}.

\begin{figure}
\begin{center}
\includegraphics[width=8cm]{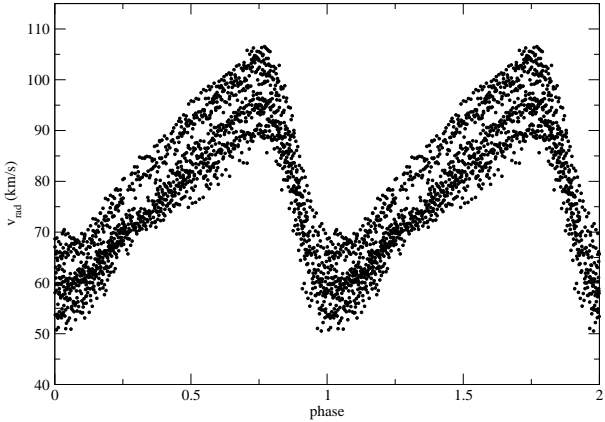}   
\end{center}
\caption[]{\label{rsgrurv1} Radial velocities of RS~Gru, phased with the 
pulsation period.} 
\end{figure}

To measure the orbital period of the system, we obtained medium-resolution
spectroscopy on 16 nights between 2003 and 2005 using the SSO230 instrument. In
addition, we observed the star with the AAT on 1 night in 2006. The whole
phased dataset is shown in Fig.\ \ref{rsgrurv1}, where the continuously
changing shift in the systemic velocity is evident. 

\begin{figure} 
\begin{center} 
\includegraphics[width=8cm]{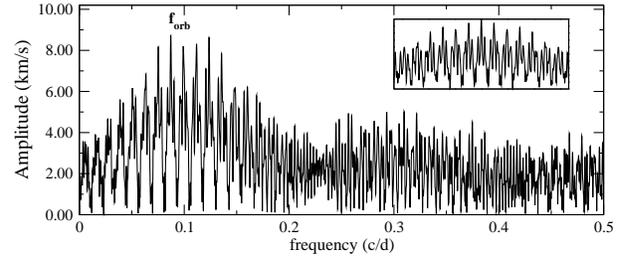}   
\end{center} 
\caption[]{\label{rsgrurvfou} Fourier spectrum of the
$\gamma$-velocities of RS~Gru. Inset shows the spectral window.} 
\end{figure}

\begin{figure}
\begin{center}
\includegraphics[width=8cm]{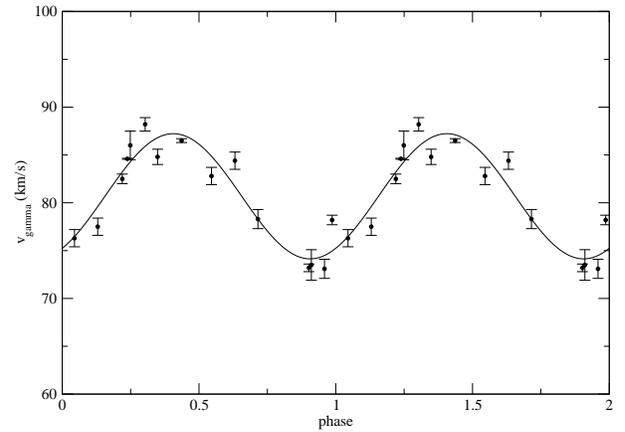}   
\end{center}
\caption[]{\label{rsgrurv2} $\gamma$-velocity variation of RS~Gru phased with
${\rm P_{orb}=11.5~d}$.} 
\end{figure}

We performed a period analysis of the radial velocities, which revealed the
main pulsation frequency at $f_{1}$=6.802~c/d, its integer harmonics ($2f_{1}$,
$3f_{1}$), and a low frequency component at about $\sim$0.11~c/d, corresponding
to a period of 9 days. However, we did not accept this as the orbital period
because the high-amplitude pulsation and random sampling may interplay and thus
render the results unreliable. Therefore, we determined the orbital period as
follows. First, we selected the best-defined single-night radial velocity curve
to fit a smooth trigonometric polynomial to the phased RV data. Then we used
the fixed polynomial to determine individual $\gamma$-velocities for each night
by fitting the zero-point only. This way we could determine the center-of-mass
velocity on 15 nights (listed in Table\ \ref{rsgrugamma}). The Fourier spectrum
of the data (Fig.\ \ref{rsgrurvfou}) shows a broad hump of peaks around 0.1~c/d
with the highest peak at 0.087~c/d, which we identify as the most likely
orbital frequency. The phased $\gamma$-velocities (Fig.\ \ref{rsgrurv2}) show a
reasonably convincing sine-wave, for which the best-fit curve (solid line in
Fig.\ \ref{rsgrurv2}) indicates a velocity amplitude of K=6.5~\kms.

 From these, we can estimate the mass of the companion from the mass function
\citep{hil01}: $f(M)=(1.0361 \times
10^{-7})(1-e^{2})^{3/2}K^{3}P=M_{2}^{3}\sin^{3}{i}/(M_{1}+M_{2})^{2}=3.3 \times
10^{-4}M_{\odot}$. Assuming that the $\delta$ Scuti mass is about
1.5--2.5~$M_{\odot}$, the calculated minimum mass for the companion at different
inclinations is shown in Table\ \ref{rsgrumass}. The derived masses show that
the companion of the RS~Gru is most likely a low-mass star. We can also
estimate the semimajor axis of the system, which is about $\sim$0.1~AU
($a=0.11~{\rm AU}$ at $M_{2}=0.09~M_{\odot}$ and $a=0.13~{\rm AU}$ at
$M_{2}=0.89~M_{\odot}$).

\begin{table}    
\begin{center} 
\caption{\label{rsgrugamma} Center of mass velocities of RS~Gru.}
\label{rsgrugamma}   
\begin{tabular}{|ccccc|}  
\hline\hline    HJD-2 400 000 & ${\rm v_{\gamma}}$ &  & HJD -2 400
000& ${\rm v_{\gamma}}$ \\  
(d) & (\kms) &   & (d) & (\kms) \\ 
\hline  
52922.0214 & 73.5$\pm$0.2 &   & 53524.2197 & 88.2$\pm$0.3 \\
53274.1377 & 82.8$\pm$0.2 &   & 53600.0627 & 73.2$\pm$0.2 \\
53275.1222 & 84.4$\pm$0.2 &   & 53601.0413 & 78.2$\pm$0.2\\
53276.0892 & 78.3$\pm$0.2 &   & 53604.0516 & 86.0$\pm$0.8 \\
53520.2614 & 73.1$\pm$0.2 &   & 53605.1980 & 84.8$\pm$0.3 \\
53521.2482 & 76.3$\pm$0.2 &   & 53606.2073 & 86.5$\pm$0.1 \\
53522.2248 & 77.5$\pm$0.2 &   & 53937.2552 & 84.6$\pm$0.05 \\
53523.2573 & 82.5$\pm$0.2 &   &            &               \\
\hline\hline  
\end{tabular}  
\end{center}   
\end{table}

\begin{table}   
\begin{center}
\caption{\label{rsgrumass} Estimated mass ($M_{2}$) for the companion of RS~Gru.}  
\label{rsgrumass}  
\begin{tabular}{|cccc|}  
\hline\hline   
 & $M_{1}=1.5~M_\odot$ & $M_{1}=2~M_\odot$ & $M_{1}=2.5~M_\odot$ \\ 
\hline
Inclination & $M_{2}$ & $M_{2}$ & $M_{2}$ \\
 $(^{\circ})$ & $(M_\odot)$ & $(M_\odot)$ & $(M_\odot)$ \\ 
\hline  
90 & 0.09 & 0.11 & 0.13 \\
70 & 0.10 & 0.12 & 0.14 \\
50 & 0.12 & 0.15 & 0.17 \\
30 & 0.20 & 0.24 & 0.27 \\
10 & 0.66 & 0.78 & 0.89 \\
\hline\hline
\end{tabular} 
\end{center}  
\end{table}

There is an interesting possibility to determine the pulsation constant, which
is most useful for pulsating stars in eclipsing binaries, where the radius of
the stars can be determined accurately \citep{jor78}. Combining Kepler's third
law and the pulsation constant formula:

\begin{equation} 
\frac{a^{3}}{P^{2}_{\rm orb}}=\frac{G}{4\pi^{2}}\left(M_{1}+M_{2}\right)
\hskip5mm {\rm and} \hskip5mm Q=P_{\rm
pul}\left(\frac{M_{1}}{R^{3}_{1}}\right)^{\frac{1}{2}} 
\end{equation}

\noindent results in (using the same units):

\begin{equation}
Q=0.1159\frac{P_{\rm pul}}{P_{\rm
orb}}\left(\frac{R_{1}}{a}\right)^{-\frac{3}{2}}\left(1+\frac{M_{2}}{M_{1}}\right)^{-\frac{1}{2}}
\end{equation}

\noindent Adopting $R_{1}=2.9\pm0.1R_{\odot}$ \citep{bal78a}, $P_{\rm
pul}=0.147$~d, $P_{\rm orb}=11.5$~d, $a=0.12\pm0.01$AU,
$M_{1}=2.0\pm0.1M_{\odot}$ \citep{rod95b}, $M_{2}=0.2\pm0.02M_{\odot}$, the
resulting $Q=0.037\pm0.013$~d is consistent with pulsations in the radial
fundamental mode. The dominant sources of error are the radius, the unknown
inclination and the orbital semi-major axis, which pose a significant
limitation at this stage. It is nevertheless reassuring that the given orbital
period and size yield a consistent picture of RS Gru being a fundamental mode
pulsator.

\subsection{RY~Leporis}

The light variation of RY~Lep (HD~38882; HIP~27400; $V$=8.2~mag; $I$=8.3~mag)
was discovered by \citet{str64} and the star was thought to be an eclipsing
binary with an unknown period for more than two decades. The SIMBAD database
still lists it as an eclipsing binary. \citet{die85} obtained five nights of
observations which revealed the real HADS nature of this star. He determined
the pulsation period as 0.2254~d and also noted small cycle-to-cycle
variations, but the data were not sufficient to draw a firm conclusion.
\citet{rod95b} determined physical parameters based on one night of $uvby\beta$
observations, covering one pulsation cycle. \citet{lan02} found aperiodic or
possibly multiperiodic variations and detected binary motion in the radial
velocities with a period more than 500~days. Finally, \citet{rod04} has shown
unambiguously the multiperiodic nature of RY~Lep (f$_{1}$=4.4416~c/d,
f$_{2}$=6.60~c/d) and its binarity was also suggested from an analysis of the
O--C diagram (details have not yet been published).

\begin{figure*} 
\begin{center}
\includegraphics[width=15cm]{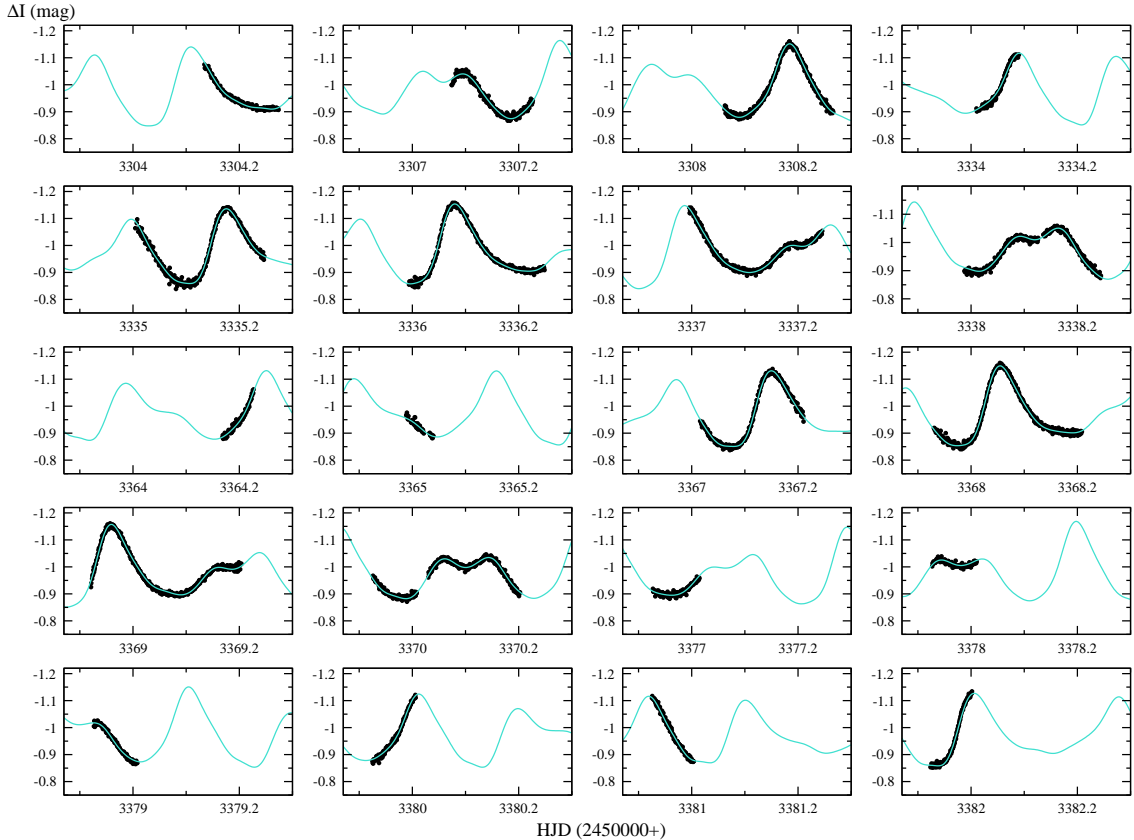}    
\end{center}
\caption[]{\label{ryleplcfit} Individual light curves of RY~Lep (small dots)
with the light curve fit (continuous line).}  
\end{figure*}

To study RY~Lep photometrically, we obtained $I$-band CCD images on 20 nights
between October 2004 and January 2005 using the APT50 instrument. We obtained
more than 5000 data points with 10~s exposures. The full log of observations is
given in Table\ \ref{obslog}. For the aperture photometry we used two
comparison stars: comp=GSC~05926-01037 ($V=9.98$~mag, $I=9.34$~mag
$B-V=1.2$~mag), check=HD~39036 ($V=8.21$~mag, $I=8.72$~mag $B-V=1.06$~mag).

\begin{figure} 
\begin{center}
\includegraphics[width=8cm]{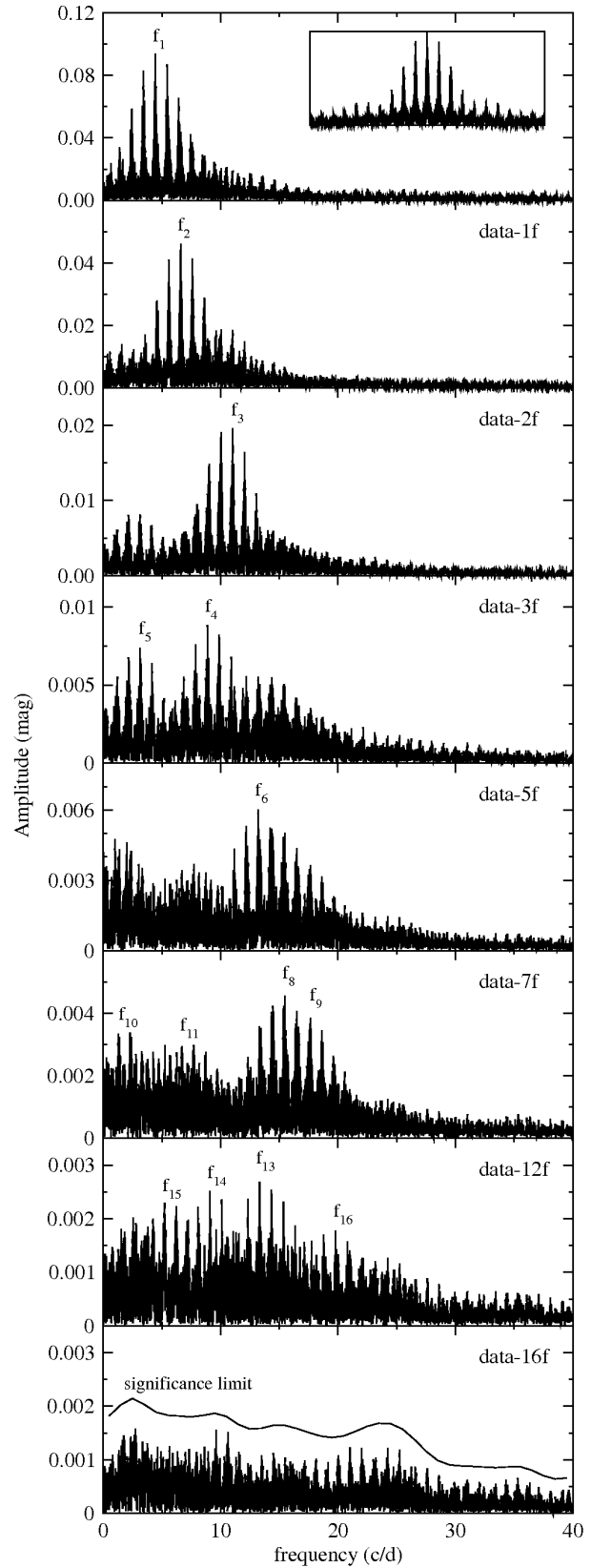}    
\end{center}
\caption[]{\label{rylepfou} Amplitude spectra of 20 nights of $I$-band data for
RY~Lep. The insert shows the window function. From top to bottom, every panel
shows an amplitude spectrum prewhitened with all frequencies marked in the
panels above. 16 frequencies can be identified with $S/N$ larger than 4.} 
\end{figure}

We performed standard Fourier analysis of the data. Fig.\ \ref{rylepfou} shows
the results of the frequency search.  We identified the two main pulsational
frequencies at f$_{1}$=4.4415~c/d and f$_{2}$=6.5987~c/d. A further 14
statistically significant peaks were found in the data, which are mainly the
various linear combinations of the two pulsation modes, and thereupon
validating the pulsational nature of f$_{2}$. The results of the period
analysis are summarized in Table\ \ref{rylepfoupar}. The final light curve fit
is shown in Fig.\ \ref{ryleplcfit}.

\begin{table}
\begin{center}
 \caption{\label{rylepfoupar} The result of the period analysis for RY~Lep.
$f_{1}$ and $f_{2}$ are the two pulsation modes, the remaining peaks are
predominantly to be the harmonics or linear combinations of these two
frequencies.}   
\label{rylepfoupar}   
\begin{tabular}{|crrrc|}   
\hline\hline
No. & Frequency & Amplitude & S/N & Frequency \\  & (${\rm d^{-1}}$) & (mmag) &  &
identification \\

\hline  
        &        &  $\pm$0.4   &   & \\  
$f_{1}$ & 4.4415 & 96.8 & 204 & \\
$f_{2}$ & 6.5987 & 46.6 & 103 & \\
$f_{3}$ & 11.0402 & 19.3 & 42 & $f_{1}+f_{2}$ \\
$f_{4}$ & 8.8830 & 10.5 & 24 & $2f_{1}$\\
$f_{5}$ & 3.1600 & 10.2 & 20 & $f_{2}-f_{1}+1.0$\\
$f_{6}$ & 13.1978 & 5.3 & 14 & $2f_{2}$\\
$f_{7}$ & 0.0046 & 5.0 & 12 & ?\\
$f_{8}$ & 15.4833 & 4.0 & 10 & $2f_{1}+f_{2}$\\
$f_{9}$ & 17.6382 & 3.7 & 10 & $f_{1}+2f_{2}$\\
$f_{10}$ & 1.3962 & 5.6 & 12 & ?\\
$f_{11}$ & 6.7237 & 3.6 & 8 & $3f_{1}-f_{2}$\\
$f_{12}$ & 0.2379 & 3.4& 8 & ?\\
$f_{13}$ & 13.3228 & 3.2 & 8 & $3f_{1}$\\
$f_{14}$ & 9.0885 & 2.9 & 8 & ?\\
$f_{15}$ & 5.2661 & 3.5 & 8 & ?\\
$f_{16}$ & 19.7970 & 1.9& 5 & $3f_{2}$\\

\hline\hline   
\end{tabular} 
\end{center}  
\end{table}

The resulting frequencies ($f_{1}$, $f_{2}$) are in very good agreement with
those by \citet{rod04}. We have two low-amplitude peaks that seem to be
significant and may be related to pulsations ($f_{14}$ and $f_{15}$), similarly
to V743 Cen, AI Vel and VW Ari, where 3, 4 and 7 frequencies were detected,
respectively \citep{mca79,wal92, liu96}. Three low-frequency components that
are presumably artifacts ($f_{7}$, $f_{10}$ and $f_{12}$). The frequency ratio
of $f_{1}$ and $f_{2}$ is 0.6731 which is not compatible with the usual
scenario of fundamental and first overtone radial modes (${\rm FU}/{\rm
1O}\approx0.77$). \citet{rod04} identified $f_{1}$ with the fundamental mode
and $f_{2}$ with a non-radial $p_{2}$ mode. The frequency ratio could also
indicate first and third overtone radial modes, for which theoretical models
predict ${\rm 1O}/{\rm 3O}\approx0.68$ \citep{san01}, but the physical
parameters of the star, such as temperature, luminosity and evolutionary mass,
are not compatible with that possibility \citep{rod95b,rod04}.

 Looking at the pulsational amplitudes of RY Lep, one can notice some
interesting features. In our data the amplitude ratio of the $f_{2}$ and
$f_{1}$ frequencies is about 0.5. Contrary to this, observations by
\citet{rod04} implied a significantly lower amplitude ratio of about 0.1. If we
use a trasformation factor F$\approx$1.7 between the $I$-band and the $V$-band
amplitudes (see fig. 2 of \citet{bal99}), numbers in Table\ \ref{rylepfoupar}
imply $\Delta V(f_{1})=164.6$~mmag and $\Delta V(f_{2})=79.2$~mmag. In
comparison, the data in \citet{rod04}, obtained between 1998 and 2002, revealed
$\Delta V(f_{1})=164.8$~mmag and $\Delta V(f_{2})=11.1$~mmag (Rodr\'iguez,
2008, personal communication). We conclude that $f_{1}$ seems to be very stable
in amplitude, whereas $f_{2}$ shows strong amplitude variations, with recent
data implying an 8-times larger amplitude.

The first spectroscopic measurement of RY~Lep was presented by \citet{pop66},
where the spectral type was determined as F0. Recently, \citet{lan02} found
clear evidence for binary motion using radial velocity measurements but the
data did not allow them to determine the orbital period which appeared to be
longer than 500 days. 

\begin{figure}
\begin{center}
\includegraphics[width=8cm]{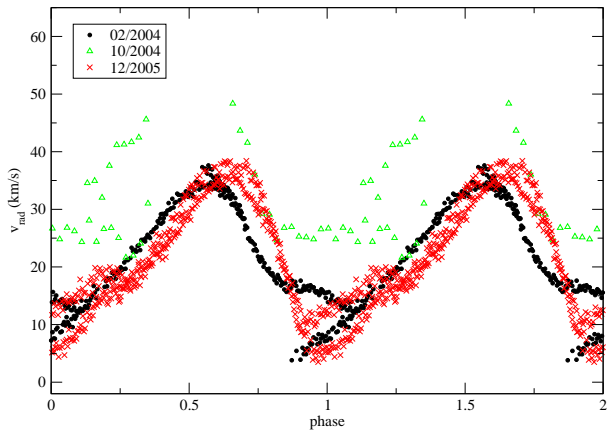}   
\end{center}
\caption[]{\label{ryleprv} The RV curve of RY~Lep phased with P=0.225~d. The
data of 6 nights in 2004 and 2005 clearly show the $\sim 25~$\kms\ of
$\gamma$-velocity  shift between the two sets of observations. Note that the
October 2004 dataset is binned.} 
\end{figure}

We obtained radial velocity measurements on four nights in 2004 and two in
2005. The RV curves of the seasonal datasets (February 2004, October 2004 and
December 2005) phased with the main period ($f_{1}$) are presented in Fig.\
\ref{ryleprv} with three different symbols.

The pulsation amplitude of RY~Lep is about 30~\kms\ and its multiperiodic
nature causes cycle-to-cycle variations in the RV curves. The February 2004 and
December 2005 datasets have basically the same $\gamma$-velocity values but the
October 2004 dataset has a higher value by about 25-30~\kms. This leads to the
conclusion that the orbital motion is clearly detected in the almost 700~day
long dataset, which is in agreement with \citet{lan02} but still does not allow
us to determine the orbital period. To estimate the approximate nature of the
companion, we assumed that the orbital period is about 730~d \citep{lan03}
and took the full range of 25~\kms\ in $v_{\gamma}$ as an estimate of the
2K$_{\rm 1}$ velocity amplitude. Repeating the same calculations as for RS Gru,
the companion's mass is about 1.1$\pm$0.15~M$_{\odot}$ and the orbital
semi-major axis is  2.3~AU, i.e. the companion is comparable to RY Lep in
mass. The lack of noticeable spectral lines from the secondary may suggest a
white dwarf but a firm conclusion would require spectra with broader coverage.

\subsection{AD Canis Minoris}

One of the best studied HADS is AD~CMi (HD~64191; HIP~38473; $V$=9.38~mag),
whose light variation was discovered by \citet{hof34} and classified as an
eclipsing binary by \citet{zes50}. The first detailed study of the star was
done by \citet{abh59} who took photometric and spectroscopic observations but
the data were not sufficient to determine the radius using the Wesselink
method. Further observations were obtained by \citet{and60,eps73, dea77,
bal83}. \citet{bre75} used {\it uvby$\beta$} photometry to determine radius,
mass and variations of physical parameters during the pulsation cycle, while
\citet{mcn85} found that the rotational velocity is smaller than 20~\kms.
Fourier decomposition of the light curve \citep{ant86} showed a surprisingly
high $\phi_{21}$ value, suggestive of overtone pulsation. However, the star
seems to pulsate in fundamental mode as other monoperiodic HADS stars do
\citep{kil93}. There has been no explanation for this phenomenon. \citet{kim90}
and \citet{kim94} determined the radius of AD~CMi, using the visual surface
brightness method and found a very good agreement with angular diameters from
theoretical and empirical relationships. \citet{kil93} analyzed 8 nights of
$UBVR$ photometry and calculated physical parameters that agree well with
\citet{bre75} and suggested the star is lying on the cool-edge of the
instability strip of the Population I stars. \citet{jia87} reported a
continuous  period increase at the star, which was confirmed by
\citet{rod88,rod90a}. The stability of light curve was studied by \citet{rod99}
who found no significant long-term changes in amplitude. The first suggestion
for binarity for AD~CMi was presented by \citet{fu96}, who found a possible
orbital period of 30 years from the O--C diagram. Most recently, \citet{hur07}
and \citet{kho07} have studied the period variations of AD~CMi using published
and new data. They deduced the presence of light-time effect due to binarity
and a slow period increase due to evolutionary effect. In addition,
\citet{kho07} detected an extra low-frequency component in the photometric
data, which provides a possible explanation for the large scatter of the O$-$C
diagram.

\begin{figure}
\begin{center}
\includegraphics[width=8cm]{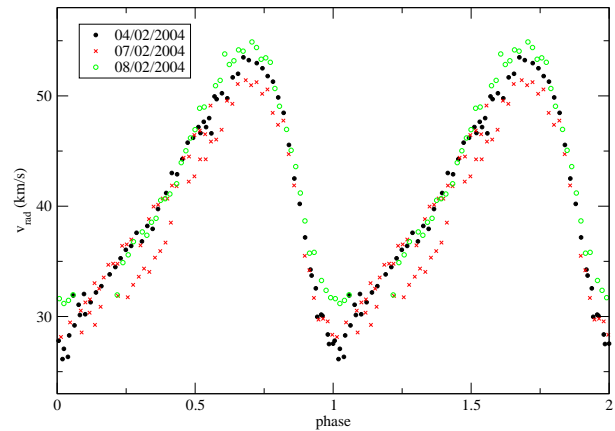}   
\end{center}
\caption[]{\label{adcmi} RV curve of AD~CMi phased with the pulsation period
(E$_{0}$=2449401.1320~d; P=0.12297443~d).} 
\end{figure}

We performed spectroscopic measurements on 3 nights in February 2004 (see
Table\ \ref{obslog}). The phased RV data (Fig.\ \ref{adcmi}) have a mean
amplitude of $\sim$25~\kms, while showing significant cycle-to-cycle variation.
The mean velocity is about 40~\kms. Two radial velocity measurements are
available in the literature \citep{abh59,bal83}. They determined
$\gamma$-velocities at 34.5~\kms\ and 38.8~\kms, respectively. \citet{hur07}
interpreted the current $O-C$ diagram as a combination of the a continuous
period increase and light-time effect. The amplitude of the orbital motion is
expected to be about 1.1~\kms\ \citep{hur07}. Considering this orbital
amplitude, our data are in good agreement with \citet{bal83}. The \citet{abh59}
data, six points in total, are of lower quality and have poor phase coverage,
so that the larger difference is still compatible with our result.

\citet{pet98} were the first to notice that AD CMi may be peculiar in terms of
luminosity because the Hipparcos parallax indicated that the star, among five
others, was situated approximately 3 mag below the standard P--L relation. They
even suggested the possible existence of an ``AD CMi group''. We have checked
the new reduction of the Hipparcos data \citep{van07}. The updated parallax
$\pi=6.20\pm1.47$~mas differs by about 1-$\sigma$ from the original value at
$\pi_{\rm old}=8.40\pm1.73$~mas. While the new value pushes the absolute
magnitude of  AD CMi about 1~mag brighter, there is still a significant shift
left unexplained. Studies of HADS/SX Phe variables in clusters and nearby
galaxies \citep[e.g.][]{por06} do not indicate this large spread in absolute
magnitude, so that we suspect that there might be a yet-to-identify source of
systematic error in some of the Hipparcos HADSs.

Our data also shows cycle-to-cycle variations in the shape of RV curve that are
within a range of a 2-3~\kms\ as shown in the phase diagram in Fig.\
\ref{adcmi}.  This might be due to the presence of an additional pulsation
mode but our data are not extensive enough to resolve multiple modes. If this
secondary mode is the same one reported by \citet{kho07}, its amplitude must
change in time, because the very low amplitude in the \citet{kho07} data is
hardly compatible with the $2-3$~\kms\ cycle-to-cycle RV change we find in the
spectroscopic measurements. It is interesting to add that the \citet{abh59}
data showed a larger peak-to-peak amplitude of about 35~\kms, which may also be
due to cycle-to-cycle variations caused by a second excited mode.

\subsection{BQ~Indi}

\begin{figure*} 
\begin{center}
\includegraphics[width=15cm]{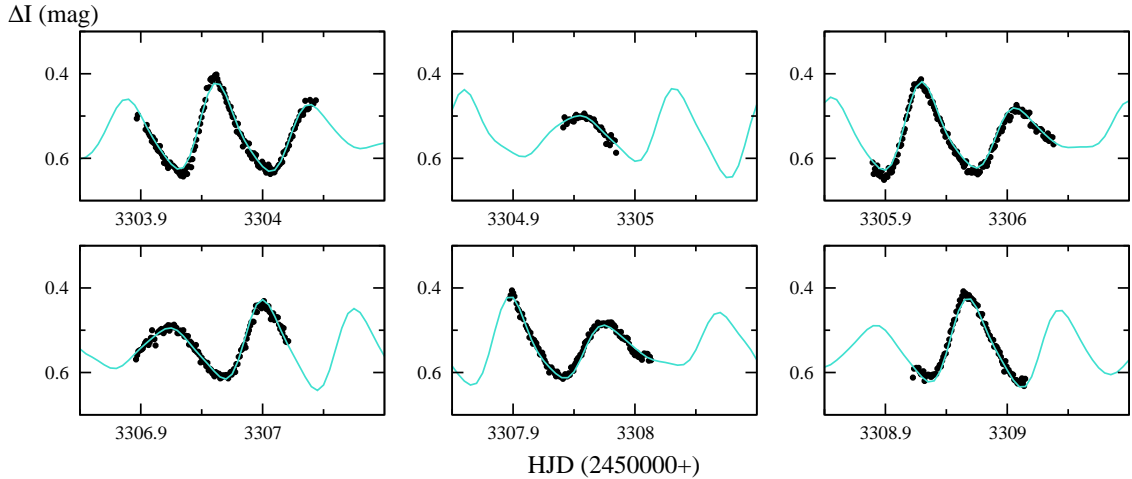}    
\end{center}
\caption[]{\label{bqind} Individual light curves of BQ~Ind (small  dots) with
the five-component fit.}
\end{figure*}

BQ~Ind (HD~198830; HIP~103290) was discovered to be a variable by the Hipparcos
satellite and has a mean magnitude $V$=9.8~mag, $I$=9.7~mag and a period of
0.0819877~d \citep{per97a}. The multiperiodic nature of the star first
discovered by \citet{ste03}, who determined two frequencies
($f_{1}$=12.1951~c/d, $f_{2}$=15.7686~c/d), corresponding to the fundamental
and first overtone modes. Since then, no further observations have been
reported in the literature.

We performed CCD photometry on BQ~Ind on six consecutive nights in 2004 with
APT50. More than 700 data points were obtained with 30-40~s exposure time in
$I$-band; a log of observations is given in Table\ \ref{obslog}. For the
aperture photometry we used two comparison stars: comp=GSC~08800-00069
($V=10.6$~mag, $I=9.51$~mag, $B-V=1.28$~mag), check=PPM~774605 ($V=10.5$~mag,
$I=9.34$~mag, $B-V=1.38$~mag).

\begin{figure} 
\begin{center}
\includegraphics[width=8cm]{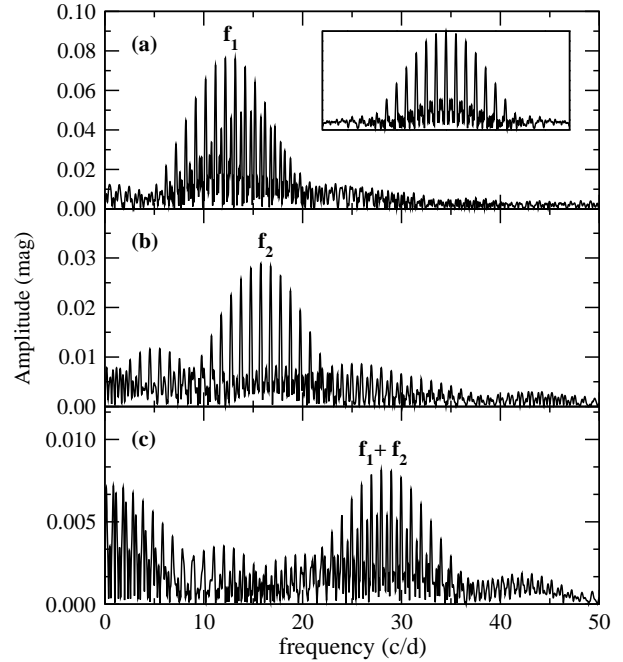}    
\end{center}
\caption[]{\label{bqindfou} Fourier analysis of BQ~Ind. Panel {\bf a:}
Amplitude spectrum of the complete dataset. The insert shows the window
function. Panel {\bf b:}  After removal of the main period and its harmonics,
the secondary period is clearly seen. Panel {\bf c:} After removal of the
secondary period, the next peak is the linear combination of the two
frequencies.}
\end{figure}

The amplitude spectrum is shown in Fig.\ \ref{bqindfou}. The primary peak was
found at $f_{1}=12.1961~d^{-1}$ and the next prewhitening step yielded the
secondary frequency at $f_{2}=15.7593~d^{-1}$. After subtracting the two main
frequencies we ended up at their linear combination and then the integer
harmonics ($2f_{1}$, $3f_{1}$) of the primary frequency. Their parameters are
summarized in Table\ \ref{bqindfoupar}.

The resulting two frequencies ($f_{1}$, $f_{2}$) confirm the double-mode nature
of BQ~Ind, and are in very good agreement (within 1\%) with the frequencies
determined by \citet{ste03}, with no further frequencies in the residuals. The
period ratio is $f_{1}/f_{2}=0.7739$, which suggests fundamental ($f_{1}$) and
first-overtone ($f_{2}$) mode pulsation. The Fourier-fit of the individual
light curves is presented in Fig.\ \ref{bqind}.

\begin{table}   
\begin{center}
\caption{\label{bqindfoupar} The result of the period analysis for BQ~Ind.}  
\label{bqindfoupar}  
\begin{tabular}{|crrrc|}  
\hline\hline   
No. & Frequency & Amplitude & S/N & Mode\\ 
& (${\rm d^{-1}}$) & (mmag) & &combination\\
\hline  
        &         & $\pm$1.1  &   & \\
$f_{1}$ & 12.1961 & 71.1 & 54 & \\
$f_{2}$ & 15.7593 & 30.2 & 29 & \\
$f_{3}$ & 27.9580 & 9.0 & 6 & $f_{1}+f_{2}$\\
$f_{4}$ & 24.3903 & 10.4 & 7 & $2f_{1}$\\
$f_{5}$ & 36.5671 & 2.9 & 5 & $3f_{1}$\\
\hline\hline  
\end{tabular} 
\end{center}  
\end{table}

\subsection{ZZ~Microscopii}

The short-period variability of ZZ~Mic (HD~199757; HIP~103684) was discovered
by \citet{chu61}. Its average $V$ magnitude is 9.43~mag and the pulsation
period is  0.0654~d. The first detailed analysis of this star was done by
\citet{leu68}, who found cycle-to-cycle variation in ultraviolet light and also
detected a period decrease. Later the photoelectric observations and data
analysis \citep{cha71,rod99} did not confirm any change in the light curve
shape. \citet{per76} reanalyzed Leung's observations and deduced two periods:
0.0654~d and 0.0513~d, suggesting fundamental and first overtone pulsations
\citep{bal78b}. Previously, \citet{bes69} analyzed spectrophotometric and
spectroscopic observations and determined the pulsation constant, masses and
absolute magnitudes, concluding the first-overtone pulsating nature of ZZ~Mic.
The first radius determination of the star was carried out by \citet{bal78b}.
The last analysis of the star was done by \citet{rod99}, who studied the
stability of the light curve and did not find any significant long-term
amplitude change.

\begin{figure} 
\begin{center} 
\includegraphics[width=8cm]{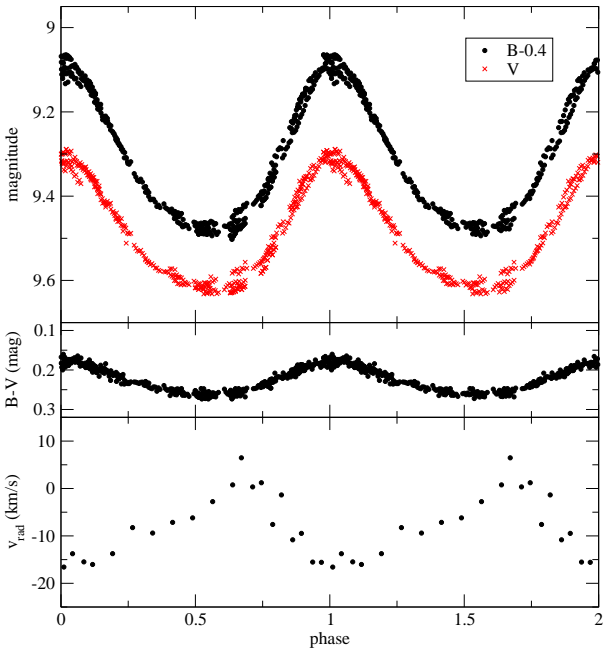}   
\end{center} 
\caption[]{\label{zzmicfaz} Standard light, colour and radial velocity
variations  of ZZ~Mic (E$_{0}$=2453305.9819~d; P=0.0671835~d).}   
\end{figure}

We took three nights of photoelectric observations in 2004 using $B$, $V$
filters on SSO60. For the calculations of differential magnitudes we used the
following two comparison stars: comp=HD~199639 ($V=7.28$~mag, $B-V=0.16$~mag)
and check=HD~200320 ($V=8.96$~mag, $B-V=0.51$~mag). The phase diagrams are
shown in Fig.\ \ref{zzmicfaz}.

\begin{figure}
\begin{center}
\includegraphics[width=8cm]{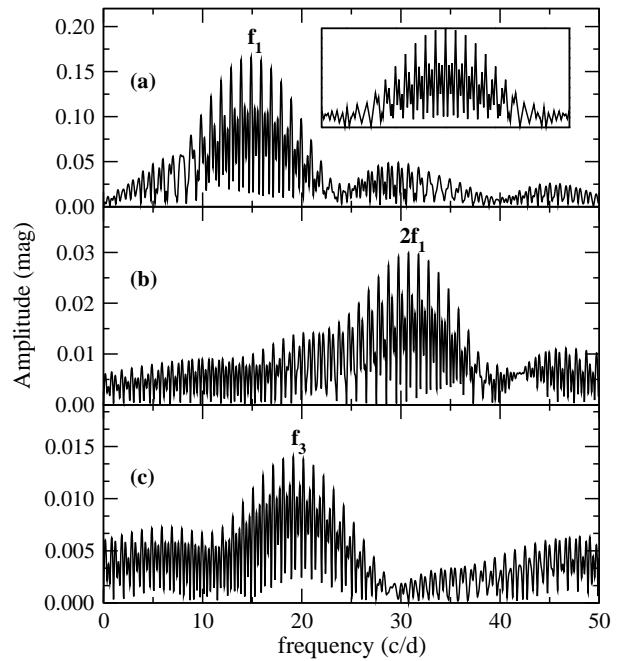}   
\end{center}
\caption[]{\label{zzmicfou} Fourier spectra of ZZ~Mic with the pre-whitening
steps. The insert shows the spectral window.} 
\end{figure}

Since the discovery of ZZ~Mic, it has been controversial in terms of changing
light curve shape and being multiperiodic. In order to study the question, we
performed a period analysis of our admittedly meagre  V-band data. The
pre-whitening steps are plotted in Fig.\ \ref{zzmicfou}, while the resulting
parameters are listed in Table\ \ref{zzmicfoupar}. The Fourier spectrum is
dominated by the main pulsational period ($f_{1}=$14.896~c/d) and its harmonic.
With a much lower amplitude (A$_{3}=$14~mmag compared to A$_{1}=$147.3~mmag) we
detected a secondary period  at $f_{3}=$19.15~c/d which is a reasonably good
agreement with that of \citet{per76} ($f_{1}=$15.3~c/d, $f_{2}=$19.5~c/d). The
$S/N$ ratio of this frequency is 8, which is quite low compared to $f_{1}$ and
2$f_{1}$ but still above the significance limit. 

\begin{figure} 
\begin{center}
\includegraphics[width=8cm]{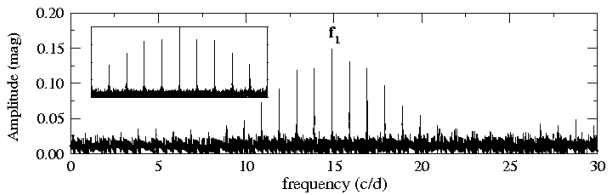}    
\end{center}
\caption[]{\label{zzmicasas} The Fourier spectrum of ZZ~Mic using the ASAS
data. The insert shows the spectral window.}  
\end{figure}

Because of the limited data we have, we tried to detect the secondary frequency
in other publicly available data. We analyzed the data from the All Sky
Automated Survey (ASAS) project \citep{poj02}. The Fourier spectrum of this
data (Fig.\ \ref{zzmicasas}) clearly shows the main pulsational period at
$f_{1}=$14.885~c/d but the noise level in the dataset ($\sim$23~mmag) is too
high compared to the amplitude of the secondary frequency ($\sim$14~mmag),
which prevents any detection in the ASAS data.

\begin{table}   
\begin{center}
\caption{\label{zzmicfoupar} The result of the period analysis for the ZZ~Mic.}  
\label{zzmicfoupar}  
\begin{tabular}{|crrrc|}  
\hline\hline   
No. & Frequency & Amplitude & S/N & Mode\\ 
& (${\rm d^{-1}}$) & (mmag) & &combination\\
\hline  
        &        & $\pm$1.4   &  & \\
$f_{1}$ & 14.896 & 147.3 & 85 & \\
$f_{2}$ & 30.77 & 29.7 & 42 & $2f_{1}$\\
$f_{3}$ & 19.15 & 14 & 8 & \\
\hline\hline  
\end{tabular} 
\end{center}  
\end{table}

If we accept $f_{3}$ as an independent mode, the ratio of the two modes is
$f_{1}/f_{3}=0.778$. This suggests $f_{1}$ is the fundamental mode and $f_{3}$
is the first overtone mode. The fundamental mode identification for $f_{1}$ is
strongly supported by the $ubvy\beta$ photometry of \citet{rod00}. Moreover,
$f_{1}$ must be a radial mode which was suggested from the phase shifts in $BV$
photometry by \citet{rod96}.

Moreover, a period ratio of 0.778 seems to be too large for a normal Pop.
I HADS \citep{por05, pet96} which suggests that ZZ Mic is a Pop. II star.
However, the value of $f_{3}$ is not too reliable, so the period ratio might be
slightly different. Studies on metal abundances and space motions \citep{bre80}
suggest that ZZ Mic is a normal Pop. I HADS, which is also supported by
\citet{rod00}.

We obtained spectra simultaneously with $BV$ light curves on one night using
SSO230. The resulted RV curve is shown in the bottom panel of Fig.\
\ref{zzmicfaz}, which is the first radial velocity curve obtained of ZZ Mic. The
full amplitude of the RV curve is 22~\kms.

\subsection{CY Aquarii}

\begin{figure}
\begin{center}
\includegraphics[width=8cm]{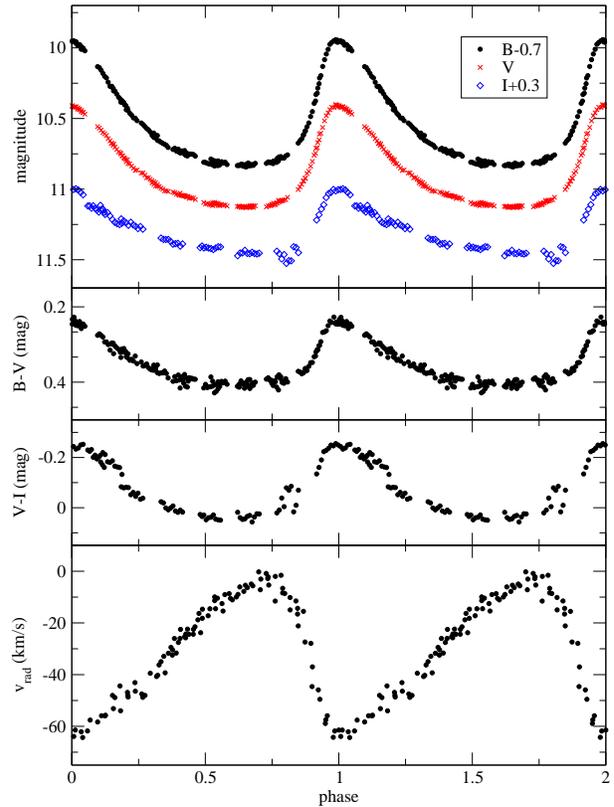}   
\end{center}
\caption[]{\label{cyaqr} Standard light, colour and radial velocity variations 
of CY~Aqr (E$_{0}$=2452920.9223~d; P=0.061038328~d).}  
\end{figure}

CY~Aqr (HIP~111719; $V$=10.7~mag, $I$=10.3~mag) is one of the shortest period
HADS in the galactic field, with a pulsation period of 0.061038d, and has been
subject to many investigations. It was discovered by \citet{hof35}. A number of
early studies on the star are listed by \citet{har61}, who estimated physical
parameters and found that the shape of the light curve varies. The period
stability was studied by \citet{ash54}, who found no change in period but
noticed a phase jump that seemed to be attributed to the star. Further studies
were made by \citet{zis68,nat72,boh80}. Changes in light curve shape and the
possibility of another period were also investigated in several papers, e.g.
\citet{els72, fit73,fig78} but both phenomena were discounted later by
\citet{gey75,per75,pur84,hin97}. Finally, \citet{coa94} set a definite upper
limit of 1.5~mmag in $V$ for the amplitude of any long-lived secondary period.

\citet{kam85} made a thorough period study and his results indicated the
presence of random fluctuations in pulsation frequency that cannot be explained
by considering only evolution. Other period change studies were performed by
\citet{rol86,mah88,pow95}. \citet{mcn96} determined physical properties.
\citet{fu94} suggested that the period changes due to the presence of an unseen
companion with an orbital period of around 50~years. \citet{zho99} and
\citet{fu03} studied the O--C diagram to characterize long-term period
evolution. They found a long-term cyclic component, and both suggested possible
binarity for CY~Aqr with an orbital period of $\sim$62.4~yr and $\sim$52.5~yr,
respectively.

We obtained 3 nights of standard $BVI$ photoelectric, 5 nights $I$-band and 5
nights CCD $V$-band photometry with SSO60, APT50 and P60 between 2003 and 2007.
The integration time was 15s with SSO60 and 50s with the APT50. The full log of
observations is given in Table\ \ref{obslog}. Differential magnitudes were
calculated using the following comparison stars: comp=GSC~00567-02242
($V=9.8$~mag, $I=8.96$~mag, $B-V=1.38$~mag) and check=GSC~00567-01242
($V=10.6$~mag, $I=9.62$~mag, $B-V=1.14$~mag). The resulting light and colour
curves are plotted in the top three panels of Fig.\ \ref{cyaqr}. We determined
new times of maximum that are listed in Table\ \ref{maxtimes}. The O--C diagram
(not shown) is in a very good agreement with light-time solution determined by
\citet{fu03}.

We obtained spectroscopic observations on two nights in 2003 and 2004. The mean
radial velocity is --38~\kms. This value is in a good agreement with previous
data by \citet{str49} and \citet{fer87}, who measured --32~\kms\ and --40~\kms,
respectively. The predicted amplitude of mean velocity change due to the
binarity is $\sim$1.4~\kms\ \citep{zho99} which is comparable to the accuracy
of our observation. Furthermore, due to the high eccentricity and long period
of the binary system, the mean velocity changed only slightly in the last 10
years, which is far beyond our detection limit. In conclusion, our radial
velocity observations do not contradict the current understanding of the nature
of CY~Aqr.

\subsection{BE~Lyncis}

We obtained high-resolution spectroscopy of BE~Lyn on three nights with MH150
in order to detect possible binarity or additional pulsational frequencies. The
period change of this star inspired a series of studies by our group
\citep{kis95,der03,sza08} and, while the initial orbital elements of the
suspected binary system were ruled out and hence leaving the binarity
unconfirmed, the lack of spectroscopic data in the literature has kept this
star in our focus.

\begin{figure}
\begin{center}
\includegraphics[width=8cm]{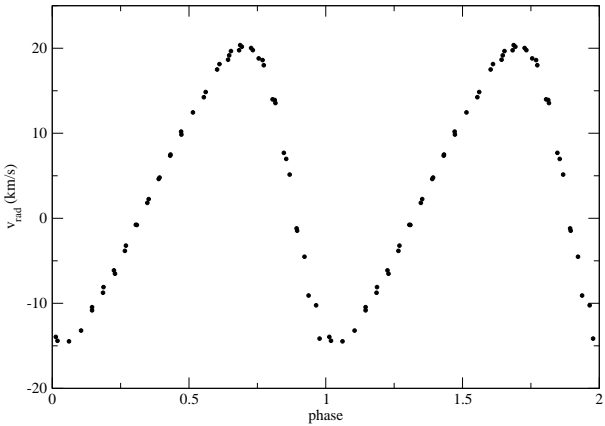}   
\end{center}
\caption[]{\label{belynrvfaz} Phased RV curve of BE~Lyn (HJD=2449749.4651~d,
P=0.09586952~d).} 
\end{figure}

To our knowledge, our radial velocity measurements are the first obtained for
BE Lyn. The phased RV curve is shown in Fig.\ \ref{belynrvfaz}, where we see
characteristic shape and velocity amplitude for fundamental mode pulsation. The
center-of-mass velocity is measured at 3.4~\kms, while the amplitude of the
variation is $\sim$34~\kms. There is no sign of gamma velocity change during
the three nights of observation and we also could not detect any non-radial
mode pulsation in the RV curve. Further study of the data in terms of velocity
gradient within the stellar atmosphere is in progress.

\subsection{Period updates for XX~Cygni, DY~Pegasi and DY~Herculis}

We performed times-series photometry on XX~Cyg, DY~Peg and DY~Her. Previous
studies of these stars are listed in \citet{der03}. Since then, only DY Peg was
studied by \citet{hin04}, who explained the period change of the star with two
period breaks rather than continuously decreasing rate, as was previously
thought.

We obtained 3 nights of $V$-band CCD photometry on XX Cyg with P60 and Sz40 and
1-1 night on DY Her and DY Peg with P60 during 2007 and 2008. The journal of
observations is given in Table\ \ref{obslog}. We determined new times of
maximum that are listed in Table\ \ref{maxtimes}. The updated O--C diagrams
contain these and recently published data collected from the literature
\citep{age03,hub05a,hub05b,bir06,hub06,kli06,hub07a,hub07b}.

\begin{figure}
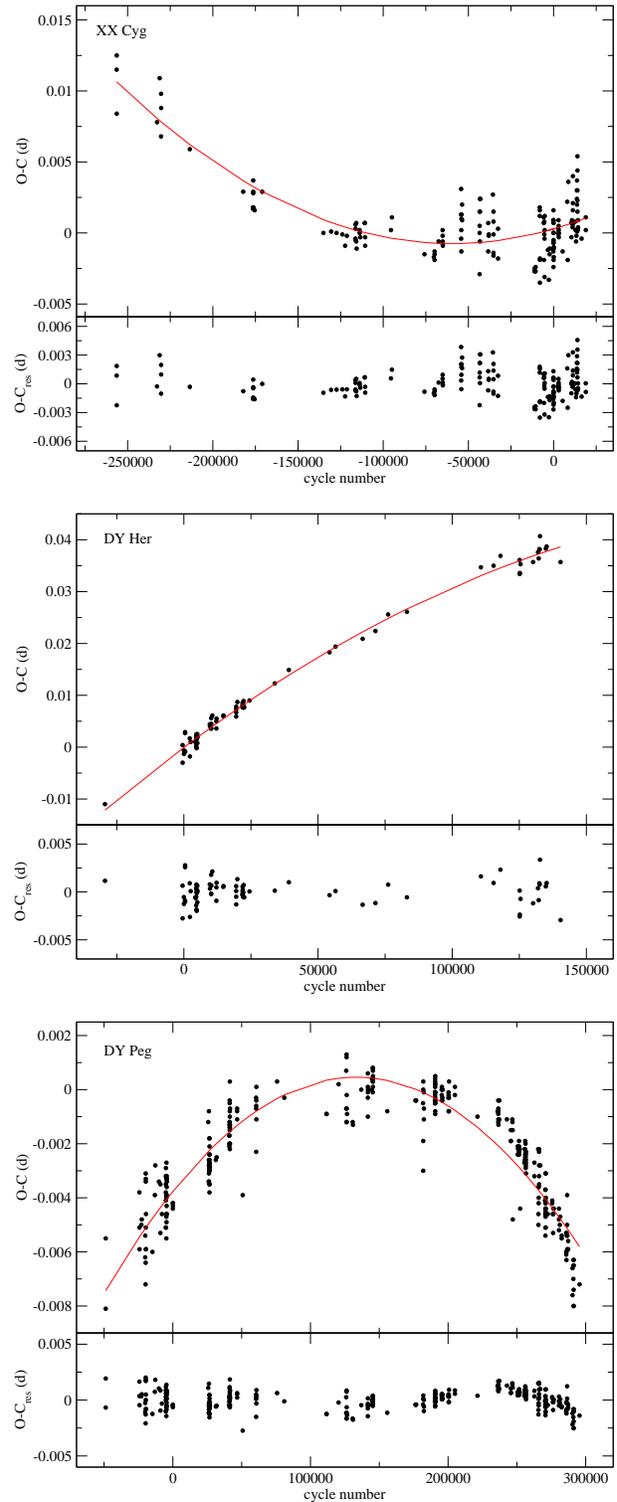

\begin{center}
\includegraphics[width=8cm]{fig16a.eps}
\vskip3mm   
\includegraphics[width=8cm]{fig16b.eps}   
\vskip3mm   
\includegraphics[width=8cm]{fig16c.eps}   
\end{center}
\caption[]{\label{oc} O--C diagram of DY Peg, DY Her, XX Cyg.} 
\end{figure}


We calculated the O--C diagram of XX Cyg using the following ephemeris: ${\rm
HJD_{max}}=2451757.3984+0.13486513 \times E$ \citep{der03}, and the resulting
diagram is shown in the top panel of Fig.\ \ref{oc}. The parabolic fit form of
the O--C diagram is: ${\rm HJD_{max}}=0.0003+3.49 \times 10^{-8}E+2.93\times
10^{-13}E^{2}$ with an rms of 0.00152~d. The second-order coefficient
corresponds to a relative rate of period change $\frac{1}{P} \frac{dP}{dt}=1.17
\times 10^{-8}{\rm yr}^{-1}$, which is only slightly different from the value
of $1.13 \times 10^{-8}{\rm yr}^{-1}$ given by \citet{bla03}. Therefore, we can
conclude that the period of XX Cyg has been increasing at a same rate of the
last few years. The residuals have no signs of any other change.

The O--C diagram of DY Her was calculated with the following ephemeris: ${\rm
HJD_{max}}=2433439.4871+0.1486309 \times E$ \citep{der03}. The diagram is shown
in the middle panel of Fig.\ \ref{oc}. The O--C diagram was fitted with a
parabolic form of ${\rm HJD_{max}}=-0.001(3)+3.98 \times 10^{-7}E-8.98\times
10^{-13}E^{2}$ with an rms of 0.0013~d, which gives $\frac{1}{P}
\frac{dP}{dt}=-2.96 \times 10^{-8}{\rm yr}^{-1}$. This is 6\% difference in the
rate of period change previously given by \citet{der03}. The residuals do not
show any other period change, so we can conclude that the present results in
period change is very well agreed with the previous studies of DY Her and the
star is showing continuous slow period decrease.

Finally, we also updated the O--C diagram of DY Peg using the following
ephemeris: ${\rm HJD_{max}}=2432751.9655+0.072926302 \times E$ \citep{mah87}
and shown the final O--C diagram in the bottom panel of Fig.\ \ref{oc}. We
performed a parabolic fit of the diagram that resulted in the following form:
${\rm HJD_{max}}=-0.003(8)+6.34 \times 10^{-8}E-2.38\times 10^{-13}E^{2}$ with
an rms of 0.0008~d. From this, we derived $\frac{1}{P} \frac{dP}{dt}=-3.27
\times 10^{-8}{\rm yr}^{-1}$ which is in a good agreement with \citet{mah87},
\citet{pen87} and \citet{der03}. The residuals of the O--C diagram show some
signs of cyclic change over 100 000 cycles but the present data are
insufficient to draw a firm conclusion.

\section{Conclusions}

We have carried out multicolor photometry and medium- and high-resolution
spectroscopy of ten bright high-amplitude $\delta$ Scuti stars over 5 years.
Our aim was to detect binarity and/or multiperiodicity in HADS variables in
order to deepen our knowledge of interaction between oscillations and binarity.

To put our binary targets in a broader context, we have compiled a complete
list of binary HADS variables, presented in Table\ \ref{hadscmass}. How do RS
Gru and RY Lep, the two newly confirmed spectroscopic binaries, compare with
other known systems? Looking at the 8 stars in Table\ \ref{hadscmass}, we can
see three distinct groups with markedly different orbital periods. RS~Gru is
one of the shortest-period binary and it is interesting to note that neither
UNSW-V-500 nor RS~Gru show evidence of multimode pulsations. RY~Lep is similar
to SZ~Lyn both in the orbital period and the reasonably large mass of the
companion. These intermediate-period systems are also promising for detecting
spectral features of the companion in the ultraviolet or infrared region, thus
allowing a full dynamical mass determination. To be able to detect
spectroscopically the binary nature of the long-period systems, will require
very high-precision spectroscopy, since the expected v$_{\gamma}$ change is in
the range of 1~\kms.

\begin{table}   
\begin{center}
\caption{\label{hadscmass} Summary of the estimated masses of the companions in
the known HADS binary systems. Sources for $P_{\rm orb}$ and masses are: (1)
\citet{chr07}, (2) \citet{mof88}, (3) \citet{fu08} (4) \citet{fu99}, (5) 
\citet{hur07}, (6)
\citet{fu04}.}  
\label{hadscmass}  
\begin{tabular}{|lrrc|}  
\hline\hline   
Star & $P_{\rm 	orb}$ & $m_{\rm comp} (M_\odot)$ & Refs. \\
\hline  
UNSW-V-500 & 5.35~d & $\sim$0.3 & 1 \\
RS Gru & 11.5~d & 0.1--0.2 & present paper \\
RY Lep &  730~d: & $\sim$1.1 & present paper \\
SZ Lyn & 3.2~yr & 0.7--1.6 & 2 \\
KZ Hya & 26.8~yr & 0.83--3.4 & 3 \\
BS Aqr & 31.7~yr & 0.1--0.33 & 4 \\
AD CMi & 42.9~yr & 0.15--1.0 & 5 \\
CY Aqr & 52.5~yr & 0.1--0.76 & 6 \\
\hline\hline
\end{tabular} 
\end{center}  
\end{table}

To summarize, the main results of this paper are the follows:

\begin{enumerate}

\item We monitored RS~Gru spectroscopically on 17 nights in order to measure
the orbital period. We derived the orbital period as 11.5~days.

\item We confirmed the multimode pulsation of RY~Lep from CCD photometry,
detecting and refining the frequencies of two independent modes. Spectroscopic
measurements also show the multimode pulsation. We detected the orbital motion
in the radial velocity curve, confirming the preliminary results of
\citet{lan02} on the binary nature of RY~Lep. Our 700~day-long dataset is in
good qualitative agreement with \citet{lan02}. The limits on the orbital period
and RV amplitude suggest a binary companion of about 1~M$_{\odot}$, possibly a
white dwarf star.

\item The radial velocity curve of AD~CMi shows cycle-to-cycle variations that
support the presence of a low frequency mode pulsation reported by
\citet{kho07}. The center-of-mass velocity is in good agreement with the
previous measurement by \citet{bal83} and does not contradict the binary
hypothesis of the star, since the predicted $\gamma$-velocity change is around
1~\kms.

\item We obtained the first spectroscopic measurements for BE~Lyn. The RV curve
has an amplitude of $\sim$34~\kms\ and the center-of-mass velocity is 3.4~\kms.

\item We confirmed the double-mode nature of BQ~Ind, corresponding to the
fundamental and first overtone modes.

\item We detected a low-amplitude secondary period in the photometry of ZZ~Mic
but further observations are needed to confirm its validity. The RV curve has a
full amplitude of 22~\kms.

\item We updated the O--C diagram for CY~Aqr, corroborating binarity found by
\citet{zho99} and \citet{fu03}. Our radial velocity data are in a good
agreement with previous observations by \citet{str49} and \citet{fer87} but
have better accuracy.

\item We obtained new time series photometry on XX Cyg, DY Her and DY Peg and
updated their O--C diagrams with new times of maximum. DY Her and DY Peg show
continuous period decrease, while XX Cyg has continuous period increase.

\end{enumerate}

\noindent Further analysis of the photometric and spectroscopic observations
(e.g. determination of physical parameters) will be presented in a subsequent
paper.

\section*{Acknowledgments} 

We are grateful to the referee for the comments that improved the paper. AD is
supported by an University of Sydney Postgraduate Award. GyMSz is supported by
the Bolyai J\'anos Research Fellowship of the Hungarian Academy of Sciences.
The research was supported by the Hungarian OTKA Grant T042509. This work has
been supported by the Australian Research Council. The authors are grateful for
Prof. Chris Tinney for taking the high-quality service observations with the
AAT. We thank the CfA TAC for their support by providing telescope time for
this project. The NASA ADS Abstract Service was used to access data and
references. This research has made use of the SIMBAD database, operated at
CDS-Strasbourg, France.


\appendix

\section{Full log of observations.}

\begin{table*}   
\begin{center}
\caption{\label{obslog} Journal of observations.}  
\label{obslog}  
\begin{tabular}{|lcrrclcrrc|}  
\hline\hline   
Date & Filter & Instrument & Data points & Obs. length & Date & Filter &
Instrument & Data points & Obs. length \\
\hline  
\bf{RS Gru} & & & & & \bf{AD CMi} & & & &\\
2003--10--07 & $B,V,I$ & SSO60 & 93 & 3.6~h &    2004--02--04 & $spec.$ & SSO230 & 63 & 4.9~h \\
2003--10--09 & $B.V,I$ & SSO60 & 145 & 4.4~h &	 2004--02--07 & $spec.$ & SSO230 & 69 & 4.3~h \\
2003--10--11 & $B,V,I$ & SSO60 & 125 & 2.9~h &	 2004--02--08 & $spec.$ & SSO230 & 47 & 2.6~h \\
2003--10--12 & $B,V,I$ & SSO60 & 121 & 2.9~h &   \bf{BQ Ind} &&&&\\
2004--10--02 & $B,V,I$ & SSO60 & 141 & 3.6~h &   2004--10--25 & $I$ & APT60 & 153 & 3.5~h \\
2003--10--09 & $spec.$ & SSO230 & 151 & 3.9~h &  2004--10--26 & $I$ & APT60 & 46 & 1.1~h \\
2004--09--25 & $spec.$ & SSO230 & 146 & 3.3~h &  2004--10--27 & $I$ & APT60 & 156 & 3.6~h \\
2004--09--26 & $spec.$ & SSO230 & 170 & 3.4~h &  2004--10--28 & $I$ & APT60 & 133 & 3.0~h \\
2004--09--27 & $spec.$ & SSO230 & 99 & 2.3~h &	 2004--10--29 & $I$ & APT60 & 120 & 2.8~h \\
2004--09--28 & $spec.$ & SSO230 & 14 & 0.3~h &	 2004--10--30 & $I$ & APT60 & 98 & 2.2~h \\
2005--05--28 & $spec.$ & SSO230 & 14 & 0.7~h &   \bf{ZZ Mic} &&&&\\
2005--05--29 & $spec.$ & SSO230 & 179 & 3.8~h &  2004--10--27 & $B,V$ & SSO60 & 148 & 2.7~h \\
2005--05--30 & $spec.$ & SSO230 & 203 & 4.1~h &  2004--10--29 & $B,V$ & SSO60 & 105 & 1.7~h \\
2005--05--31 & $spec.$ & SSO230 & 161 & 4.2~h &  2004--10--30 & $B,V$ & SSO60 & 124 & 2.1~h \\
2005--06--01 & $spec.$ & SSO230 & 86 & 3.4~h &	 2004--10--27 & $spec.$ & SSO230 & 41 & 2.4~h \\
2005--06--02 & $spec.$ & SSO230 & 57 & 1.4~h &	 \bf{CY Aqr} &&&&\\
2005--08--17 & $spec.$ & SSO230 & 116 & 2.9~h &  2003--10--08 & $V,I$ & SSO60 & 193 & 4.0~h \\
2005--08--18 & $spec.$ & SSO230 & 125 & 3.2~h &  2003--10--10 & $V,I$ & SSO60 & 85 & 1.7~h \\
2005--08--21 & $spec.$ & SSO230 & 8 & 1.1~h &	 2003--10--13 & $B,V$ & SSO60 & 175 & 2.9~h \\
2005--08--22 & $spec.$ & SSO230 & 37 & 3.5~h &	 2004--11--24 & $I$ & APT50 & 43 & 1.3~h \\
2005--08--23 & $spec.$ & SSO230 & 98 & 3.4~h &	 2004--11--25 & $I$ & APT50 & 47 & 1.3~h \\
2006--07--21 & $spec.$ & AAT & 40 & 4.2~h &	 2004--11--26 & $I$ & APT50 & 33 & 0.9~h \\
\bf{RY Lep} & & & & &				 2004--11--27 & $I$ & APT50 & 45 & 1.3~h\\
2004--10--25 & $I$ & APT50 & 190 & 3.3~h &	 2004--11--28 & $I$ & APT50 & 49 & 1.3~h\\
2004--10--28 & $I$ & APT50 & 246 & 3.7~h &	 2007--07--25 & $V$ & P60 & 101 & 1.2~h\\
2004--10--29 & $I$ & APT50 & 335 & 4.9~h &	 2007--07--26 & $V$ & P60 & 103 & 1.8~h\\
2004--11--24 & $I$ & APT50 & 105 & 1.9~h &	 2003--10--08 & $spec.$ & SSO230 & 87 & 3.6~h \\
2004--11--25 & $I$ & APT50 & 366 & 5.8~h &       2004--07--04 & $spec.$ & SSO230 & 22 & 1.6~h \\
2004--11--26 & $I$ & APT50 & 345 & 6.1~h &	 \bf{BE Lyn} &&&&\\
2004--11--27 & $I$ & APT50 & 385 & 6.0~h &       2007--10--25 & $spec$ & MH150 & 10 & 1.0~h \\
2004--11--28 & $I$ & APT50 & 380 & 6.2~h &       2007--10--26 & $spec$ & MH150 & 22 & 2.1~h \\
2004--12--24 & $I$ & APT50 & 98 & 1.4~h &	 2007--10--27 & $spec$ & MH150 & 20 & 2.0~h \\
2004--12--25 & $I$ & APT50 & 60 & 1.2~h &	 \bf{XX~Cyg} & & & & \\ 
2004--12--27 & $I$ & APT50 & 324 & 4.7~h &       2007--07--25 & $V$ & P60 & 117 & 3.2~h \\
2004--12--28 & $I$ & APT50 & 464 & 6.7~h &	 2007--07--27 & $V$ & P60 & 201 & 3.6~h \\
2004--12--29 & $I$ & APT50 & 471 & 6.8~h &	 2008--07--29 & $V$ & Sz40 & 85 & 3.3~h \\
2004--12--30 & $I$ & APT50 & 419 & 6.6~h &	 \bf{DY~Her} & & & & \\
2005--01--06 & $I$ & APT50 & 144 & 2.1~h &	 2007--07--22 & $V$ & P60 & 131 & 2.2~h \\
2005--01--07 & $I$ & APT50 & 140 & 2.0~h &	 \bf{DY~Peg} & & & & \\ 
2005--01--08 & $I$ & APT50 & 137 & 1.9~h &	 2007--07--23 & $V$ & P60 & 145 & 2.7~h \\
2005--01--09 & $I$ & APT50 & 135 & 1.9~h &\\	 
2005--01--10 & $I$ & APT50 & 130 & 1.9~h &\\
2005--01--11 & $I$ & APT50 & 128 & 1.9~h &\\
2004--02--05 & $spec.$ & SSO230 & 178 & 4.5~h &\\ 
2004--02--06 & $spec.$ & SSO230 & 134 & 3.9~h & \\
2004--10--26 & $spec.$ & SSO230 & 28 & 1.3~h & \\
2004--10--27 & $spec.$ & SSO230 & 75 & 3.9~h & \\
2005--12--19 & $spec.$ & SSO230 & 233 & 5.5~h & \\
2005--12--20 & $spec.$ & SSO230 & 203 & 5.5~h & \\

\hline\hline
\end{tabular} 
\end{center}  
\end{table*}

\end{document}